\documentclass{aa}
\usepackage{txfonts}
\usepackage{graphicx}
\usepackage{float}
\usepackage{natbib}
\bibpunct{(}{)}{;}{a}{}{,}

\begin{document}
\titlerunning{MIR-interferometry of NGC~2264~IRS~1}
\authorrunning{R.Grellmann et al.}
\title{Mid-infrared interferometry of the massive young stellar object NGC~2264~IRS~1\thanks{Based on observations collected at the European Organisation for Astronomical Research in the Southern Hemisphere, Chile, observing programs 076.C-0725(B) and 082.C-899(A).}}
\author{R. Grellmann\inst{1} \and Th. Ratzka\inst{1} \and S. Kraus\inst{2} \and H. Linz\inst{3} \and Th. Preibisch\inst{1} \and G. Weigelt\inst{4}}
\institute{Universit\"ats-Sternwarte, Ludwig-Maximilians-Universit\"at M\"unchen, Scheinerstr. 1, 81679 M\"unchen, Germany
\and University of Michigan, Department of Astronomy, Ann Arbor, MI 48109-1090, USA
\and Max-Planck-Institut f\"ur Astronomie, K\"onigstuhl 17, 69117 Heidelberg, Germany
\and Max-Planck-Institut f\"ur Radioastronomie, Auf dem H\"ugel 69, 53121 Bonn, Germany}
\date{Received / Accepted}
\abstract{The optically invisible infrared-source NGC~2264~IRS~1 lying north of the Cone Nebula is thought to be a massive young stellar object
($\sim 10\,M_{\sun}$). Although strong infrared excess clearly shows that the central object 
is surrounded by large amounts of circumstellar material, no information about the
spatial distribution of this circumstellar material has been available until now.}
{We used the ESO Very Large Telescope Interferometer to perform long-baseline interferometric 
observations of NGC~2264~IRS~1  in the mid-infrared regime. 
Our observations resolve the circumstellar material around NGC~2264~IRS~1,      
provide the first direct measurement of the angular size of the    
mid-infrared emission, and yield direct constraints on the spatial distribution of the dust.}
{We analyze  the spectrally dispersed interferometric data taken with MIDI at two different                     
position angles and baseline lengths. We use different approaches (a geometrical model, a temperature-gradient model,          
and radiative transfer models) to jointly model the observed interferometric visibilities and
the spectral energy distribution.}
{The derived visibility values between $\sim 0.02$ and $\sim 0.3$ show that the mid-infrared emission is clearly 
resolved. The characteristic size of the MIR-emission region is $\sim 30-60$~AU; this value is 
typcial for other YSOs with similar or somewhat lower luminosities. A comparison of the sizes for the two position angles shows a significant elongation of the dust distribution. Simple spherical envelope
models are therefore inconistent with the data.
The radiative transfer modeling of our data suggests that we observe a geometrically
thin and optically thick circumstellar disk with a mass of about $0.1\,M_{\sun}$.}
{Our modeling indicates that NGC~2264~IRS~1 is surrounded by a
flat circumstellar disk that has properties similar to disks typically found 
around lower-mass young stellar objects. This result supports the assumption
that massive young stellar objects form via accretion from circumstellar disks.}

\keywords{techniques: interferometric - stars: individual: NGC~2264~IRS~1 - stars: formation-stars: circumstellar matter}

\maketitle

\section{Introduction\label{Object}}The question of how massive stars ($> 8\,M_{\sun}$) form still remains unclear because observations of massive young stellar objects (MYSOs) are rare \citep{Zinnecker2}. This has several reasons. First of all, these 
objects are surrounded by large amounts of gas and dust, which absorb nearly all the optical and near-infrared (NIR) radiation from the central source, thus making a direct observation very difficult. Second, their large distances mean that to study 
them we need data of high angular resolution. Other difficulties are the fast evolution of MYSOs and that they are rarely found in isolation. 

Two different formation scenarios for MYSOs are currently discussed: the coalescence scenario and a formation similar to low mass protostars. In the first scenario, several low mass stars merge and finally form a single, more massive star 
\citep{Bonnell1,Zinnecker1}. In the second scenario, the radiation pressure problem that would prevent spherical accretion as soon as the protostellar mass exceeds $\sim 10\,M_{\sun}$ is solved by stellar outflows, which open a cavity 
in polar direction, through which the radiation can escape and the radiation pressure in the equatorial plane is lowered. Therefore, an infall of material in this plane is possible. Observations have provided some evidence that massive stars can 
form in an analogous way to low mass stars. For instance, circumstellar disks (e.g., IRAS~20126+4104, \citealt{Cesaroni}; AFGL~490, \citealt{Schreyer3}; IRAS~13481-6124, \citealt{Kraus1}), molecular outflows, or collimated jets 
\citep[e.g.,][]{Cunningham} are often interpreted as evidence of ongoing accretion. From the theoretical side, both the competitive accretion \citep{Bonnell2} and the turbulent core model \citep{McKee} agree with the formation of disks. 
Furthermore, simulations demonstrate that high-mass stars may form by disk accretion \citep{Kuiper}. 

Traditionally, the spatial distribution of the surrounding dust was usually studied by the modeling the spectral energy distribution (hereafter SED). Such SED model fits are highly ambiguous \citep{Thamm,Menshchikov}. 
Therefore, more information is required to determine the real geometry of the circumstellar material. To test the existence of a circumstellar disk, and thus the disk accretion scenario of massive star formation, information at high angular 
resolution (corresponding to physical scales below 100~AU) is needed. Furthermore, only radiation at long wavelengths (e.g., infrared, sub-mm) is able to escape from the dense dusty surroundings of the star. Infrared long-baseline interferometry 
can thus provide direct spatial information on the required small scales. 

NGC~2264 is a young, star-forming cluster with an age of about 3~Myr located in the Mon~OB~1 molecular cloud complex. Literature values for the distance of the cluster vary between 400~pc and 1000~pc, although here we adopt the most recent value 
of 913~pc \citep{Baxter}. NGC~2264~IRS~1, the brightest IR-source in this region, was discovered in 1972 by D.~Allen (hence is also called Allen's Source) and has no optical counterpart. \citet{Allen} pointed out that since NGC~2264~IRS~1 is 
the brightest and most luminous source in this field, it is the likely source of the radiation pressure creating the Cone Nebula. It is believed to be a young star with a mass of $9.5\,M_{\odot}$, spectral type B0-B2 \citep{Thompson}, a luminosity
 of $3.5 \cdot 10^3\,L_{\odot}$ \citep{Harvey}, and a visual extinction of 20-30~mag. However, owing to its high extinction in the optical, all values are rather uncertain and should be interpreted with care, e.g., luminosity values between 
$2.3 \cdot 10^3\,L_{\odot}$ \citep{Nakano} and $4.7 \cdot 10^3\,L_{\odot}$ \citep{Schwartz} can be found in the literature. 

NGC~2264~IRS~1 is associated with molecular outflows and dense molecular clumps, and in near-infrared images a twisted jet-like feature can be seen extending north of the source \citep{Schreyer1,Ward}. 
The morphology of the circumstellar environment is unclear. Several studies did not find any hint of an asymmetric structure. \citet{Schreyer2} were unable to detect any signs of a circumstellar disk in their molecular line study and \citet{deWit} 
do not discern any disk-like structure in their images taken in the MIR with COMICS. However, models of a spherical symmetric distribution of the circumstellar material fail to reproduce the SED of the object in its entirety and were ruled out in 
several publications \citep{Tokunaga,Harvey}. However, the spatial resolution of all these studies may not be sufficient to resolve a disk-like structure.

In this paper, we present mid-infrared interferometric observations of the MYSO NGC~2264~IRS~1 performed with the Mid-Infrared Interferometric Instrument (MIDI) at the Very Large Telescope Interferometer (VLTI). The MIDI data and their reduction 
are shown in Sect.~\ref{Observations}. Different modeling approaches (geometrical, temperature-gradient, and radiative transfer models) for the visibilities and the SED are presented in Sect.~\ref{Modelling}. 
Implications for the object and final conclusions are discussed in Sect.~\ref{Discussion}.

\section{Observations and data reduction\label{Observations}}NGC~2264~IRS~1 was observed on 2005~December~23 in the course of ESO observing programme 076.C-0725(B) (PI: S. Kraus) and on 2009~February~15 in the course of ESO programme \mbox{082.C-0899(A)} 
(PI: M. Feldt) with MIDI at the VLTI (see Table~\ref{MIDIMeasurements}). MIDI \citep{Leinert1} is a two-telescope beam-combiner working in the wavelength range from 
8 to $13\,\mu$m and using the principle of a Michelson interferometer. To measure the fringe amplitude, the optical path difference (OPD) is varied with internal delay lines. The 2005 data of NGC~2264~IRS~1 were taken with a baseline of 89.1~m in the SCI-PHOT 
mode, in which interferometric and photometric data are measured simultaneously. The grism (a combination of a grating and a prism) with a spectral resolution of $\frac{\Delta \lambda}{\lambda}=230$ was used. In the same night, two calibrators were observed 
with the same configuration. \object{HD~39425} a few hours earlier and \object{HD~61421} directly after the object. The 2009 data were taken with a baseline of 40.2~m in the HIGH-SENS mode, in which interferometric and photometric data are obtained separately. 
For this observation, the prism with a spectral resolution of $\frac{\Delta \lambda}{\lambda}=30$ was used. \object{HD~48217} and \object{HD~49161} served as calibrators.
\begin{table*}
 \caption{ \label{OtherMeasurements}Measured fluxes for NGC~2264~IRS~1}
 \begin{tabular}{c c c c}
 \hline \hline
 Wavelength & Flux & Instrument/Reference  & Aperture\\ 
  {}[$\mu$m]&  [Jy] & {} & {} \\ \hline
 2-45	& 9-670		& ISO (\cite{Sloan})& $14'' \times 20 ''$ \\ 
 1.65	& 0.9 $\pm 0.04$& 2-MASS	& $4''$ \\ 
 2.2	& 6.8 $\pm 0.016$& 2-MASS	& $4''$ \\ 
 8-13	& 60-130	& MIDI 		& $0.5''$ \\ 
 12.0	& 146 $\pm 9$	& IRAS	 	& $0.75' \times 4.5'$ \\ 
 24.5	& 330		& COMICS	& $42'' \times 32''$ \\ 
 25.0	& 324	$\pm 13$& IRAS		& $0.75' \times 4.6'$\\ 
 53.0	& 980 $\pm 50$	& \cite{Harvey} & $17''$ \\ 
 60.0	& 911		& IRAS		& $1.5' \times 4.7'$\\ 
 70.0	& 960 $\pm 290$	& \cite{Sargent}& $3'$ \\ 
 100.0	& 1560		& IRAS		& $3' \times 5'$\\ 
 100.0	& 1645 $\pm 82$	& \cite{Harvey}	& $28''$ \\ 
 175.0	& 1530 $\pm 77$	& \cite{Harvey}	& $46''$ \\ 
 350.0	& 188 $\pm 70$	& \cite{Chini}	& $30''$ \\ 
 1300	& 13 $\pm 2.5$	& \cite{Chini}	& $90''$ \\ \hline
 \end{tabular} 
\end{table*}

\begin{table*}
 \caption{ \label{MIDIMeasurements}Observation log of the MIDI measurements of NGC~2264~IRS~1}
 \begin{tabular}{c c c c}
 \hline \hline
 Date & Proj. Baseline  & PA  & Mode \\ 
 {} &  [m] & [{}$^{\circ}$] & {} \\ \hline
 12/23/2005& 89.1 (UT2-UT4)& 81.1& SCI-PHOT\&GRISM\\ 
 02/15/2009& 40.2 (UT2-UT3)& 43.9& HIGH-SENS\&PRISM\\ \hline
 \end{tabular} 
\end{table*}

\begin{figure*}
\parbox{18.5cm}{\parbox{9.0cm}{\includegraphics[width=9.0cm, angle=0]{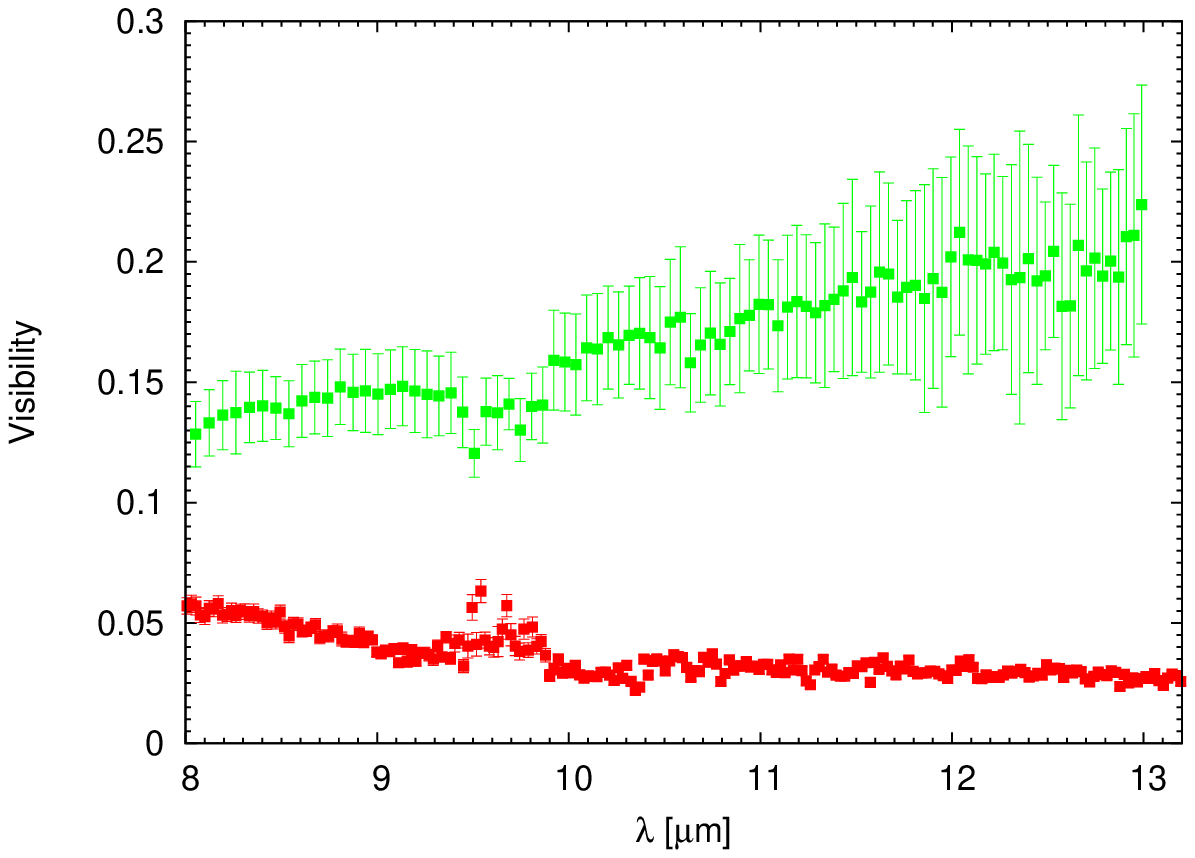}
}\parbox{9.0cm}{ \includegraphics[width=9.0cm, angle=0]{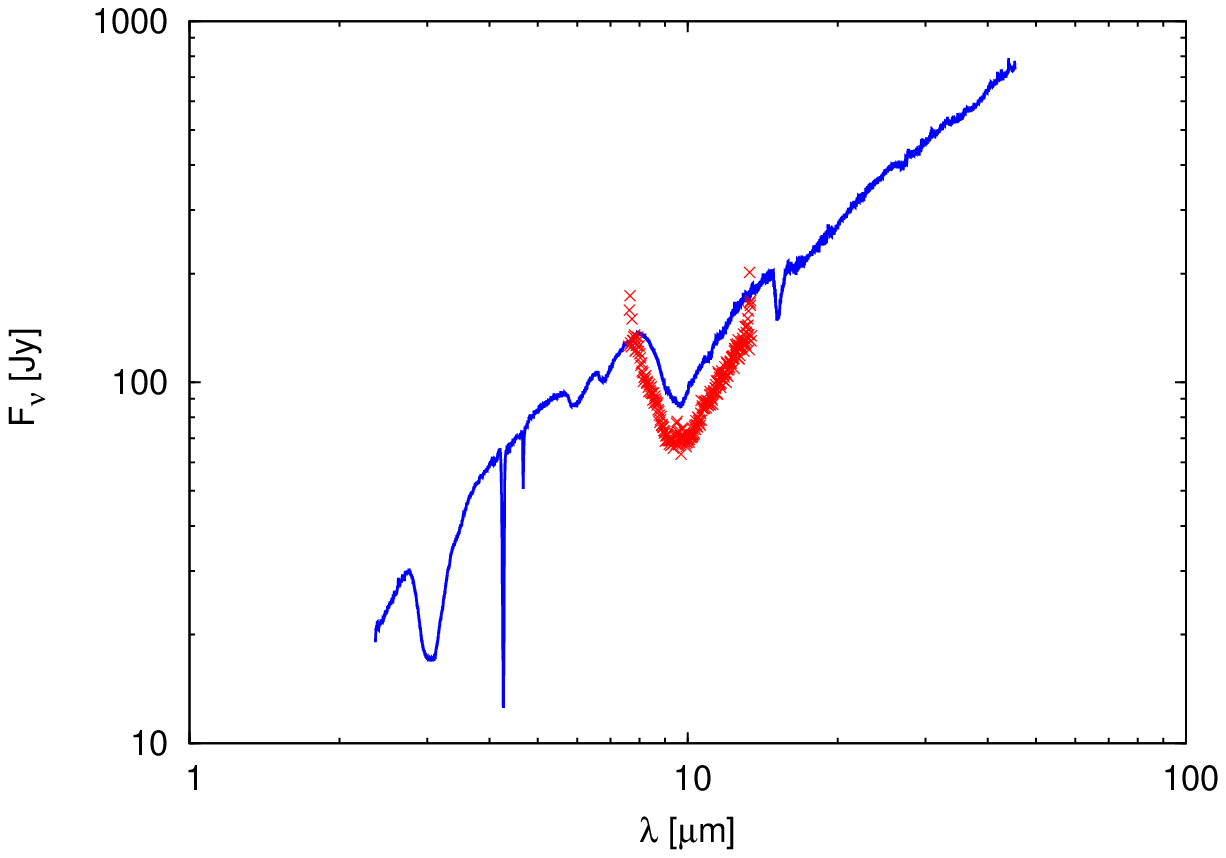}}
\caption{\label{SpektrumMIDI}\label{VisiMIDI}\textit{Left:} Visibilities of NGC~2264~IRS~1 measured with MIDI. The visibility points measured in 2005 (projected baseline length 89.1~m) are shown in red, those measured in 2009 (projected baseline 40.2~m) are shown in green. 
\textit{Right:} Spectrum of NGC~2264~IRS~1 as measured with MIDI (red crosses) in 2005 compared to the ISO spectrum (blue line).}}
\end{figure*}

For data reduction and calibration, we used the software package MIA+EWS \citep{Koehler}. With this software, the correlated flux, visibility, and the total flux can be determined. Typical errors in one single MIDI visibility measurement 
are about 10\% \citep{Leinert2}. As an estimate of the error bars, the standard deviation in the results for different calibrators was used. As the conditions during the observation in 2009 were not photometric, we did not use the 2009 photometry data, 
but the photometry from the 2005 measurement to obtain the visibility. Figure~\ref{VisiMIDI} (left) shows the calibrated visibilities for both observations.

Furthermore, a low--resolution IR spectrum of NGC~2264~IRS~1, measured with the SWS instrument aboard the ISO satellite is available (TDT No.~71602619,
PI: D. Whittet), as well as fluxes in several other wavelength bands. These data points are listed in Table~\ref{OtherMeasurements}. Fig.~\ref{SpektrumMIDI} (right) shows the total flux spectrum of NGC~2264~IRS~1 as measured by MIDI in 2005 compared 
to the total flux spectrum measured by the ISO satellite. Both spectra display a deep absorption feature over the whole MIDI wavelength range caused by silicates. The shape of both spectra (MIDI and ISO) is similar but the 
flux level of the MIDI spectrum is about 25~Jy (about 20~\%) lower, which might be due to the different beam sizes of the instruments (see Table~\ref{OtherMeasurements}). The spurious feature in the MIDI spectrum at wavelengths between 
$9.3\,\mu$m and $9.8\,\mu$m is due to atmospheric ozone. Several absorption features can be identified in the ISO spectrum \citep{Gibb}. The broad absorptions around 3.1 and $6.0\,\mu$m arise mainly from water ice, while the carrier of the $6.8\,\mu$m feature 
is more uncertain, but might be associated with solid NH$_4^+$. A rather sharp feature at $4.27\,\mu$m is commonly attributed to CO$_2$ ice \citep{Guertler}, as is the absorption seen at $15.2\,\mu$m. The feature around $4.7\,\mu$m is probably a combination of 
CO gas line absorption and a variety of ice absorption features. All these findings point to relatively large column densities of cold dense material along the line of sight towards the central source.

\section{Modeling\label{Modelling}}
We now attempt to derive the properties of NGC~2264~IRS~1 from our measurements using different modeling approaches. First, we use simple geometrical models to characterize the size of the dust emission region (see Sect.~\ref{Geometry}). To test 
for the possibility of a circumstellar disk, we compare the data with a temperature-gradient model for such a disk (Sect.~\ref{TGmodel}). To check the parameter space over a wider range, we employ the online model SED fitter by \citet{Robitaille} in Sect.~\ref{Robitaille}. 
A more detailed modeling is done in Sect.~\ref{RADMC}, where the two-dimensional radiative transfer code RADMC is used. 

\subsection{Simple geometrical models\label{Geometry}}
As a first step, one can derive an estimate of the diameter of the mid-infrared (MIR) emission region of the object from the visibilities by using simple geometrical models. For more details of this we refer to \citet{Berger} and \citet{Tristram}. 

We first consider the possibility of a binary. Its existence would result in a sinusoidal variation in the visibility (versus wavelength), ranging between a maximum and a minimum value. The maximum value depends on the extension of the MIR emission regions 
(and would be 1 for point like sources), whereas the amplitude depends 
on the flux ratio of both components (the minimum would be 0 for a flux ratio of 1:1). Owing to the slit width of MIDI (see Table~\ref{OtherMeasurements}), the maximum separation of a binary that can be identified with this instrument is $\sim 230\,\rm{AU}$. The minimum separation at which a binary 
could be resolved with the 90~m baseline is $\sim 30\,\rm{AU}$ (diffraction limit; half a period of the modulation in the visibility would be seen). Both observed visibilities show no typical sinusoidal variation, hence a binary system with a separation
 of $> 30\,\rm{AU}$ is unlikely and not considered during the modeling process. However, with our measurements we could miss very faint binary companions or systems with separations of about $30\,\rm{AU}$ oriented 
non-parallel to the position angle of the 90 m baseline.
\begin{figure}
 \includegraphics[width=9.0cm, angle=0]{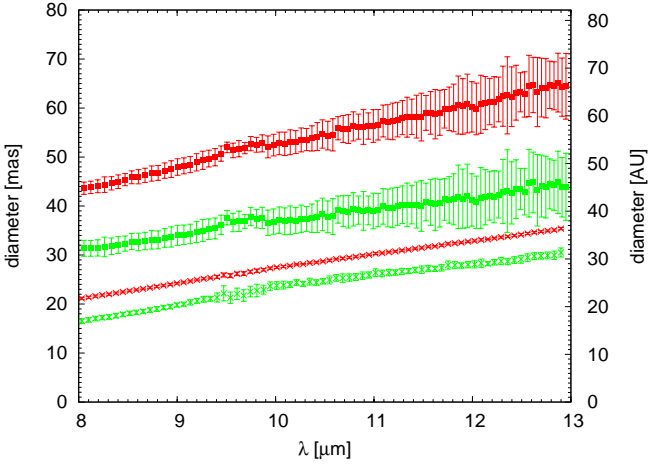}
 \caption{\label{Diameter}Mid-infrared diameter of NGC~2264~IRS~1 as a function of wavelength for the 2005 data (crosses) measured with a 90~m baseline and the 2009 data (rectangulars) measured with a 40~m baseline. The values for the diameter in AU (right y-axis) 
were computed using a distance of 913~pc. Two different models were used: a Gaussian model (red) and a uniform disk model (green). }
\end{figure}

We consider two different geometrical models for a single source: a Gaussian distribution and a uniform disk (UD). The results can be seen in Fig.~\ref{Diameter}. All curves show an increasing size of the object with increasing wavelength, as 
expected because of the different dust temperatures in different parts of the dust distribution. For the measurement performed with the 90~m baseline, the size varies between 16 and 30~mas for the Gaussian model and between 20 and 35~mas for the 
uniform disk model, whereas for the 40~m baseline, the size 
of the object lies between 30 and 44~mas for the Gaussian model and between 44 and 64~mas for the uniform disk model. As the observations were taken at different position angles, the different size estimations suggest an elongation in the 
dust distribution (e.g., a circumstellar disk or a flattened envelope). Although the intensity profile of the real source possibly differs from those distributions, the Gaussian model in particular is usually a good approximation of the 
mid-infrared morphology of YSOs. Despite all the uncertainties in the visibility measurements, hence the size estimation, the fitting of the Gaussian model still helps us identify an asymmetry with a significance of $3\,\sigma$ and the UD model within $2\,\sigma$.
This result clearly suggests that the MIR emission region is not spherically symmetric but elongated.

\subsection{Temperature gradient model\label{TGmodel}}
Many young stars of both low and high mass are known to have circumstellar disks \citep[e.g.,][]{Kraus1,Leinert2,Ratzka}. As a first approach to interpreting simultaneously SED and N-band spatial data, we therefore use a 
temperature-gradient model for a circumstellar disk as described in \citet{Malbet}. The disk specified by this model is optically thick, but geometrical thin with an inner radius 
$r_{\rm min}$ and an outer radius $r_{\rm max}$. The distribution of the temperature in the disk is described following the equation
\begin{equation}
 T(r)=T_0 {\left ( \frac{r}{r_0} \right)}^{-q},
\end{equation}
where $T_0$ is the temperature at a reference radius $r_0= 1\,\rm{AU}$. The exponent $q$ is expected to lie approximately between 0.5 for passive, irradiated, flared disks and 0.75 for passive, irradiated, flat disks \citep{Malbet}. 

One can imagine such a disk to be composed of rings with radius $r$ and thickness $dr$. If every ring emits blackbody radiation, one can evaluate the flux by integrating over the radius $r$ such that
\begin{equation}
 F_{\lambda}(0)=\frac{2\pi}{d^2}\int_{r_{\rm min}}^{r_{\rm max}} rB_{\lambda}(T(r))\cos(i) dr,
\label{MalbetSED}
\end{equation}
where $d$ is the distance to the object, $B_{\lambda}$ the Planck function for a blackbody with temperature $T(r)$, and $i$ the inclination of the disk. For the visibility, one then gets
\begin{equation}
 V_{\lambda}(b,0)=\frac{2\pi}{F_{\lambda}(0)d^2}\int_{r_{\rm min}}^{r_{\rm max}} r B_{\lambda}(T(r))J_0\left(2\pi b \frac{r}{d}\right)dr,
\end{equation}
where $J_0$ is the Bessel function of zeroth order and $b$ the baseline according to
\[b=\sqrt{u_{\Theta}^2+v_{\Theta}^2\cdot \cos^2(i)}.\]
Here, $\Theta$ is the position angle of the disk and $u_{\Theta}$ and $v_{\Theta}$ are given by
\[ u_{\Theta}=u \cos\Theta-v\sin\Theta\]
and
\[ v_{\Theta}=u \sin\Theta+v\cos\Theta.\]
We calculated a large number ($\approx 1000$) of temperature-gradient models and tried to find a model that is simultaneously able to reproduce all observations (both visibilities and SED).  For this, we varied the inner radius 
between 0.5 and 2~AU (in steps of 0.5~AU) and $T_0$ between 1400 and 3000~K in steps of 200~K. For the parameter q, we tested the values 0.5, 0.6, and 0.7. Inclination and position angle were varied in steps of $10^{\circ}$.

\begin{figure*}
\parbox{18.5cm}{\parbox{9.0cm}{\includegraphics[width=9.0cm, angle=0]{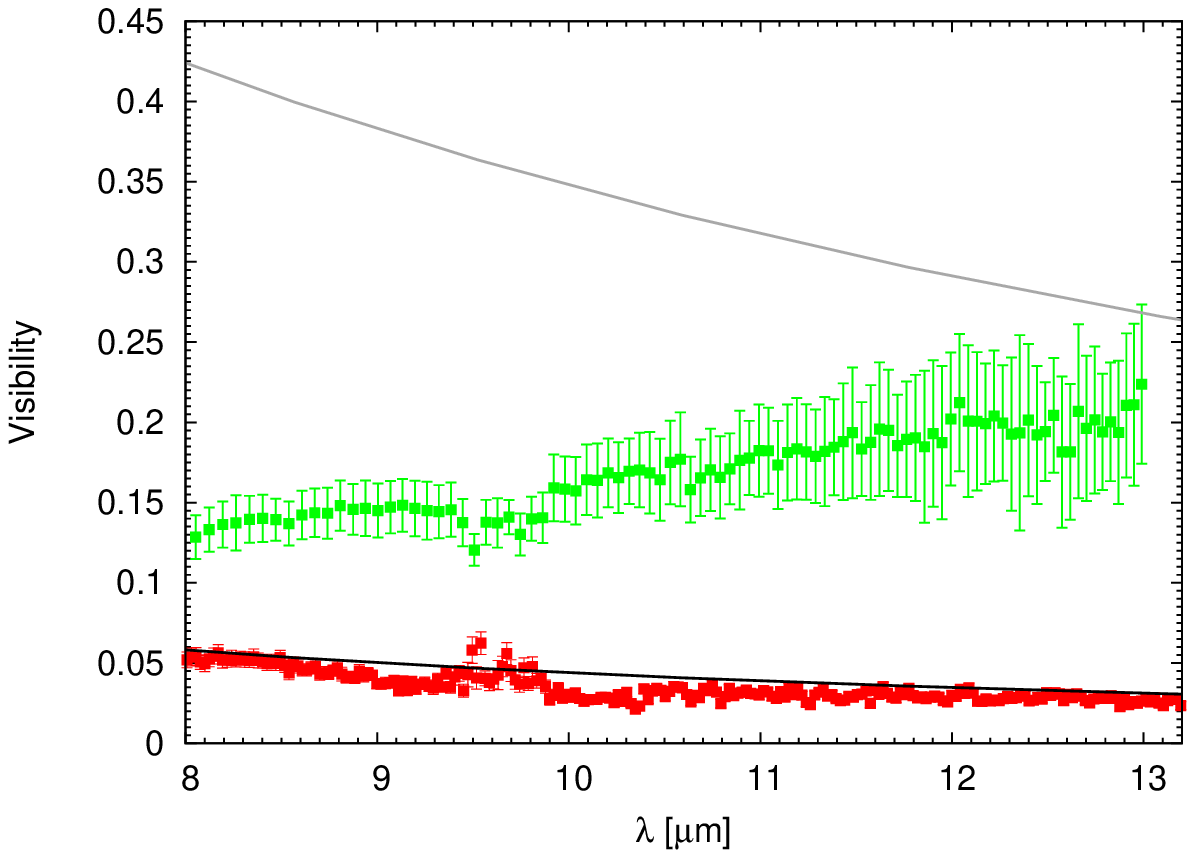}
}\parbox{9.0cm}{ \includegraphics[width=9.0cm, angle=0]{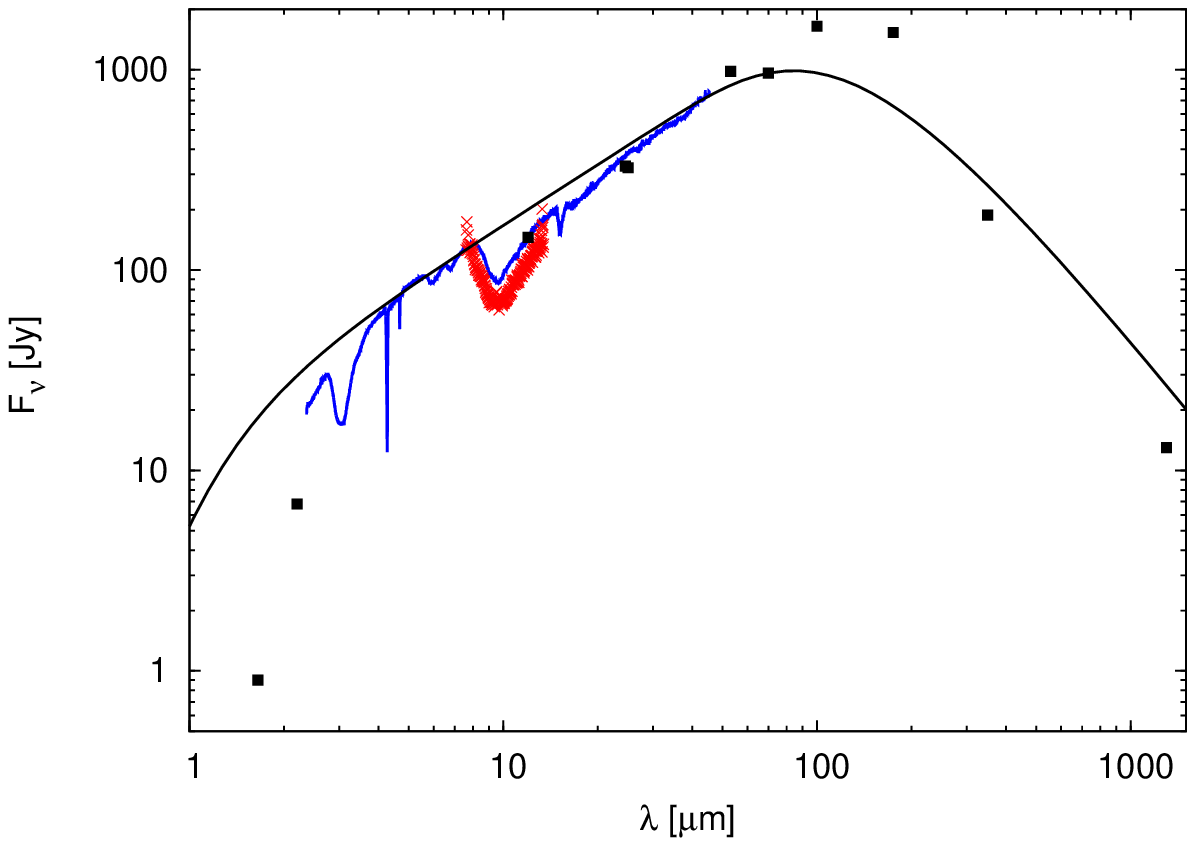}}
\caption{\label{MalbetSpektrum}\textit{Left:} Comparison of the best-fit temperature-gradient model to the visibilities of NGC~2264~IRS~1 with the model parameters described in Sect.~\ref{TGmodel} (green for the 40~m baseline, red for the 90~m baseline). 
The resulting modeled visibility for the 40~m baseline is shown by the gray line, the one for the 90~m baseline by the black line. \textit{Right:} Comparison of the SED of our best-fit temperature-gradient model (black) described 
in Sect.~\ref{TGmodel} with the ISO spectrum (blue) of NGC~2264~IRS~1. The MIDI spectrum is shown with red crosses, all other measurements from Table~\ref{OtherMeasurements} are shown with black rectangulars.}}
\end{figure*}

The temperature gradient model with the parameters listed below is able to reproduce the SED in the mid- and far-infrared, but fails in the near-infrared. All of the calculated models failed completely to reproduce the shape of the visibility 
measured with the 40~m baseline (rising values going to larger wavelengths). As the best-fit model, we therefore show the model that most successfully reproduces the other visibility curve and the SED simultaneously, although different models led to a better 
fit for the 40~m visibility at least with respect to its absolute level. The best-fit model is shown in Fig.~\ref{MalbetSpektrum} and has the following parameters:
\begin{itemize}
 \item $r_{\rm min}=0.5\,\rm{AU}$
 \item $r_{\rm max}=4000\,\rm{AU}$
 \item $T_0=2350\,\rm{K}$
 \item $i=75^{\circ}$
 \item $q=0.5$
 \item $\Theta=90^{\circ}$
\end{itemize}
The spectral features seen in the SED and the effects of the silicate absorption on the visibility cannot be reproduced, as radiative transfer effects (emission, absorption, and scattering) are not included in the model. 

A temperature of $T_0=2350\,\rm{K}$ 
at a reference radius of 1~AU would imply a temperature of 3250~K at the inner disk radius of 0.5~AU for this model. This is much too hot for the existence of normal dust grains. A possible explanation could be the existence 
of an inner gas disk \citep{Kraus2} or of dust grains with higher sublimation temperature \citep{Benisty}.

\subsection{Disk and envelope models in the Robitaille grid\label{Robitaille}}
To check a wider parameter space for different models (especially for the possibility of a circumstellar envelope), we employed the online SED fitting tool developed by \citet{Robitaille}. This tool compares 200,000 models of the SEDs of YSOs, which were precomputed 
using a 2D radiative transfer code \citep{Whitney}, with the measurements. These models are described 
by a set of 28 parameters, such as the properties of the central object (e.g., mass, temperature), the properties of the circumstellar envelope (e.g., outer radius, accretion rate, opening angle), the circumstellar disk (e.g., mass, outer disk radius, flaring), 
and the ambient density \citep[for details see][]{Robitaille1}. In addition to defining SED points, the user has to set a range for the interstellar extinction and for the distance of the object. We chose the interstellar extinction $A_V$ to be between 0 and 40~mag 
and the distance $d$ to be between 800~pc and 1000~pc. 

The number of points that can be fitted by the SED online tool is limited. Therefore, from the ISO spectrum about one flux point per one micron was used. From the MIDI spectrum, about two flux points per micron were chosen as the
shape of the spectrum changes much more in this wavelength regime because of the silicate absorption feature.

When comparing the model visibilities and fluxes to the observed data, it is
important to consider the very different (effective) beam-sizes of the
different observations.
The MIDI instrument has an effective field-of-view of $0.5''$ (slit width), corresponding to
$\approx 450$~AU at the distance of our target.
The beam sizes for all other data points beyond near-infrared wavelengths
are much larger: the beam size from which the ISO spectrum was extracted
($\approx 13\,000 \times 18\,000$~AU) is already $\approx 1100$ times larger,
and the beam sizes for all far-infrared data points are
at least $\sim 3000$ times larger than the MIDI field-of-view.
The ISO and far-infrared fluxes do therefore not only trace the
emission from the central YSO and its immediate circumstellar
material (on spatial scales of up to $\sim 500$~AU), but also contain contributions
from the surrounding extended molecular cloud, 
on spatial scales of $\ga 20\,000$~AU
($\cor \ga 0.1$~pc).

Therefore, as a first approach we include all SED points with $\lambda > 13\,\mu$m only as upper limits. To compute the visibilities, we used the HO-CHUNK code from 
\citet{Whitney} to calculate images for the model. Doing so, the ten best-fit models consist of circumstellar disks only (see Sect.~\ref{DiskRobi}). As a second approach, none of the data points was used as an upper limit, but they all were given with their errors. 
Here, rather different best-fit models are found by the fitting tool. They all consist of a large circumstellar envelope and either have no additional disk, 
or just a small ring-like structure with radii from 0.5 to 8~AU (see Sect.~\ref{EnvelopeRobi}). For similar applications of this fitting tool and its advantages and limitations in this respect, we refer to for instance \citet{Linz}, \citet{deWit2}, and 
\citet{Follert}. We note that the parameter space is not covered uniformly. In particular, the spacing of the grid parameters is much finer in the case of low-mass YSOs than for MYSOs.

\subsubsection{Robitaille disk models\label{DiskRobi}}
\begin{figure*}
\parbox{18.5cm}{\parbox{9.0cm}{\includegraphics[width=9cm, angle=0]{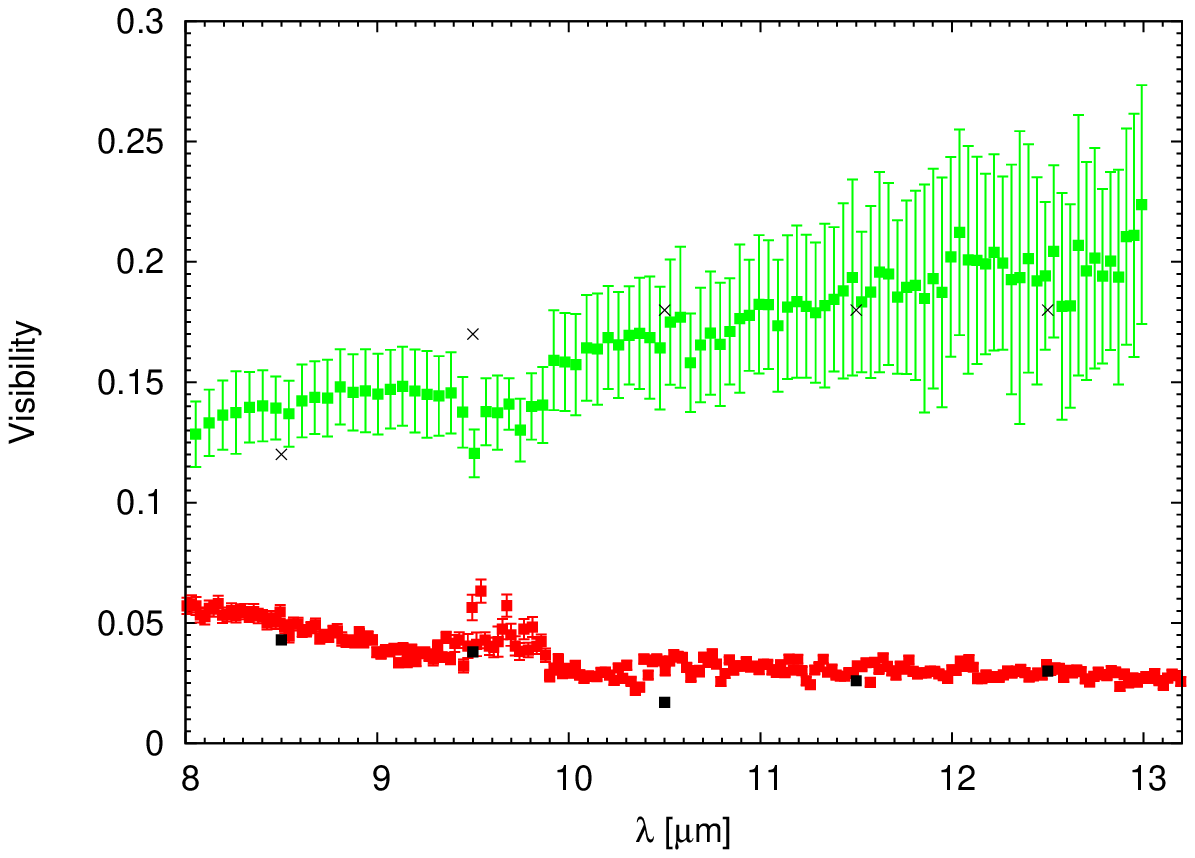}}
\parbox{9.0cm}{\includegraphics[width=9.0cm, angle=0]{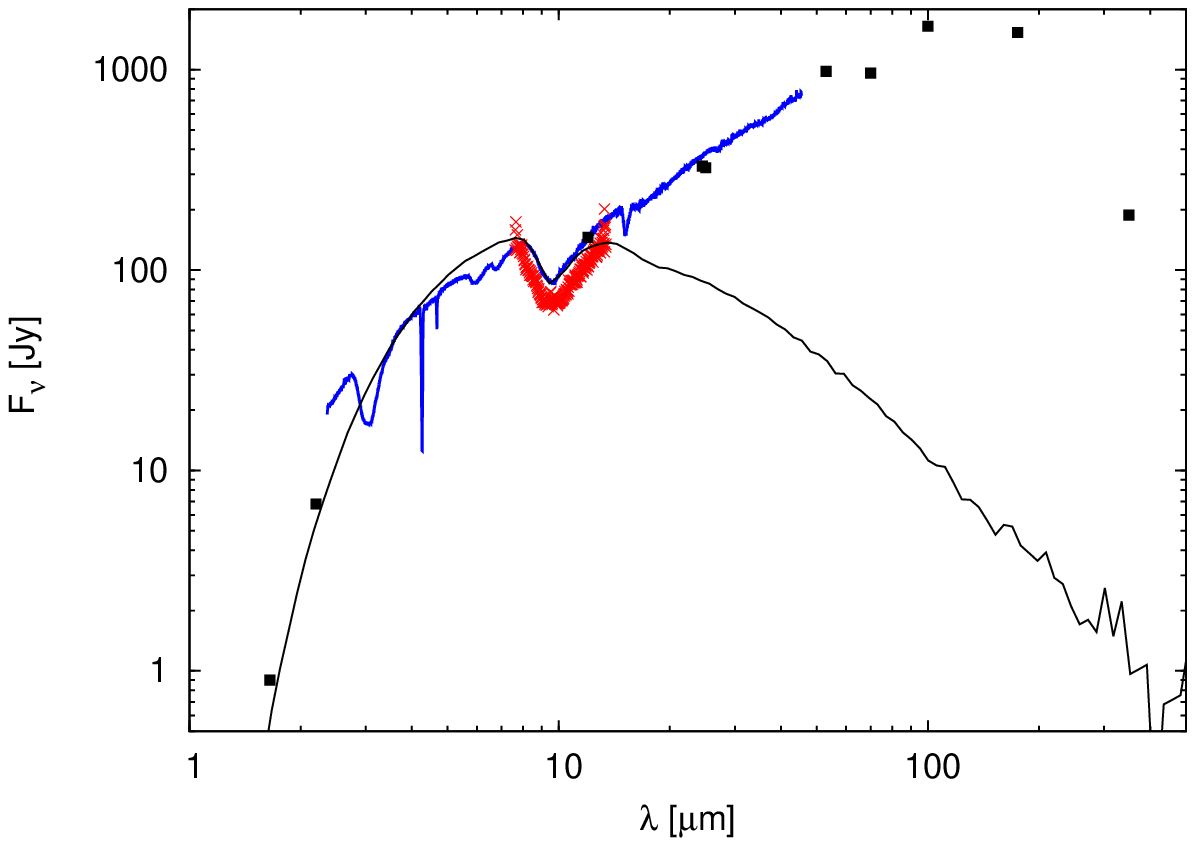}}
\caption{\label{SpektrumRobitaille}\textit{Left:} Comparison of the best-fit Robitaille disk model (black crosses 40~m baseline, black rectangulars 90~m baseline) for the visibilities of NGC~2264~IRS~1 with the MIDI visibilities (red 90~m baseline, green 40~m 
baseline). \textit{Right:} Comparison of the best-fit Robitaille model (black) for the spectrum of NGC~2264~IRS~1 with the ISO spectrum (blue), the MIDI spectrum (red crosses), and the data points from Table~\ref{OtherMeasurements} 
(black rectangulars).}}
\end{figure*}

The SED and visibilities of the best-fit Robitaille disk models are shown in Fig.~\ref{SpektrumRobitaille}. This model is able to reproduce the flux in the near- and mid-infrared. To reproduce the flux in the FIR, one would have to add for example some 
blackbody components as shown later for the RADMC model (see Sect.\ref{RADMC}). 
The model can reproduce the level of the 90~m baseline visibility, and also the 40~m baseline visibility is reproduced within the errorbars. 
The parameters of this best-fit model are (model ID~3004478):
\begin{itemize}
 \item $M_{\rm star}$ = $11.6\,M_{\odot}$
 \item $T_{\rm star}$ = 27,720~K
 \item $L_{\rm star} = 1.21 \times 10^4\,L_{\sun}$
 \item $r_{\rm out}$ = 84~AU
 \item $h_{\rm disk}\rm{(100\,AU)}$ = 5.6~AU
 \item $m_{\rm disk}$ = $0.09\,M_{\odot}$
 \item $i$ = $70^{\circ}$
 \item $A_V$ = 20.27~mag
 \item $\Theta = 40^{\circ}$
\end{itemize}
Here, $r_{\rm out}$ is the outer disk radius and $h_{\rm disk}$ would be the height of the disk when extending it to a radius of 100~AU. 

\subsubsection{Envelope plus disk models\label{EnvelopeRobi}}
\begin{figure*}
\parbox{18.5cm}{\parbox{9.0cm}{\includegraphics[width=9cm, angle=0]{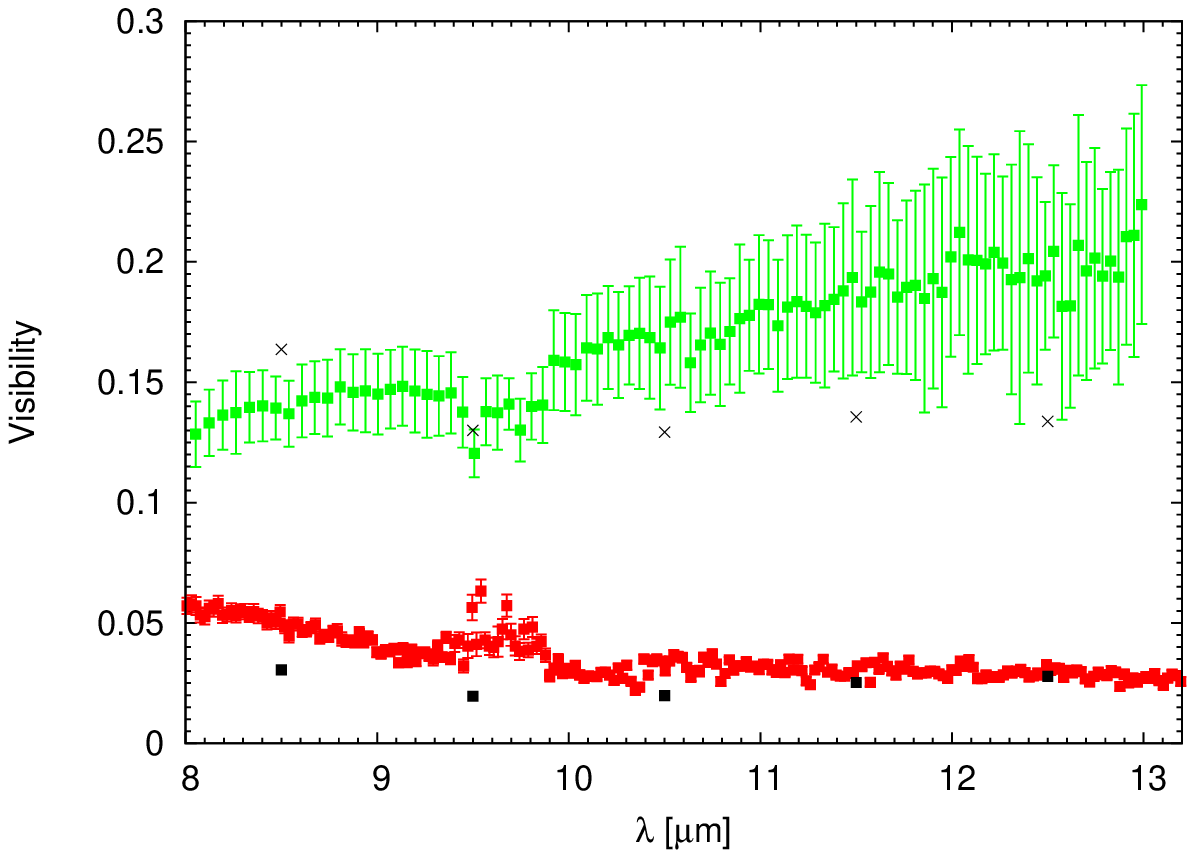}}
\parbox{9.0cm}{\includegraphics[width=9.0cm, angle=0]{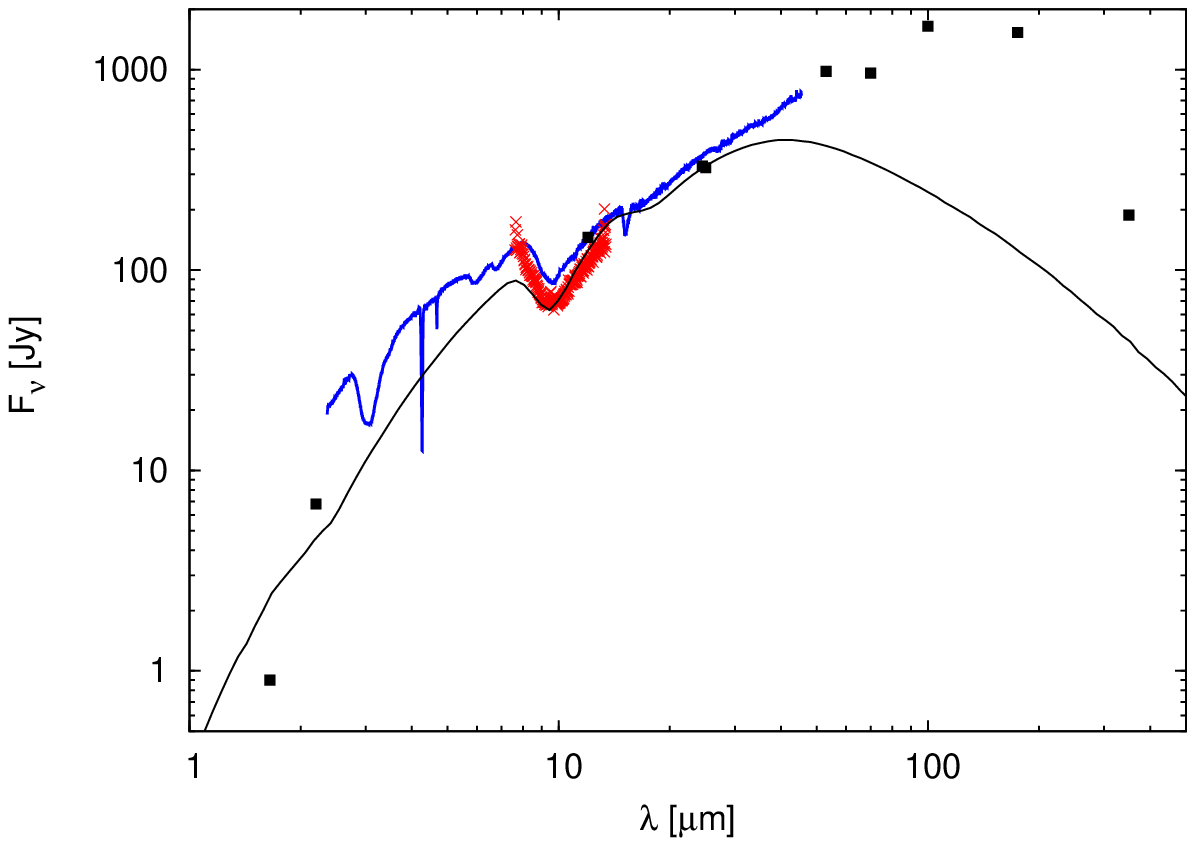}}
\caption{\label{SpektrumHuelle}\textit{Left:} Comparison of the best-fit Robitaille envelope model ( black crosses 40~m baseline, black rectangulars 90~m baseline) for the visibilities of NGC~2264~IRS~1 with the MIDI visibilities (red 90~m baseline, green 40~m 
baseline). \textit{Right:} Comparison of the best-fit Robitaille model (black) for the spectrum of NGC~2264~IRS~1 with the ISO spectrum (blue), the MIDI spectrum (red crosses), and the data points from Table~\ref{OtherMeasurements} 
(black rectangulars).}}
\end{figure*}

The Robitaille envelope model with the smallest $\chi^2$ is shown in Fig.~\ref{SpektrumHuelle}. It fails to model especially the shape of the 40~m baseline 
visibility curve. The flux in the mid-IR can be reproduced, but the model cannot reproduce the flux in the near-IR. To attain the flux level in the far-IR, one would once again have to add, e.g., a blackbody component. 
The parameters of the best-fit model are (model ID~3019090):
\begin{itemize}
 \item $M_{\rm star}$ = $8.5\,M_{\sun}$
 \item $T_{\rm star}$ = 4200~K
 \item $R_{\rm star}$ = $82\,R_{\sun}$
 \item $L_{\rm star} = 2.05 \times 10^3\,L_{\sun}$
 \item $r_{\rm envelope}$ = 36,000~AU
 \item $m_{\rm envelope} = 1.84\,M_{\sun}$
 \item $r_{\rm disk}$ = 3.6~AU
 \item $h_{\rm disk}\rm{(100\,AU)}$ = 7.7~AU
 \item $m_{\rm disk}$ = $0.2\,M_{\sun}$
 \item $i$ = $18^{\circ}$  
 \item $A_V$ = 3.8~mag.
\end{itemize}
A change in the position angle does not in this case lead to significant differences for this model, as the dust distribution is nearly symmetric. Therefore, a best-fit position angle cannot be given.
As mentioned before, this model consists mainly of a circumstellar envelope, with an outer radius of $r_{\rm envelope}$. However, there is also a disk component with a very small outer radius $r_{\rm disk} = 3.6\,\rm{AU}$. The central object is much cooler
and larger than the central object in the disk models.

\subsection{Two-dimensional radiative transfer modeling\label{RADMC}}
A more detailed physical model for a circumstellar disk is obtained by using the two-dimensional (2-D) radiative transfer model code RADMC \citep[Version 3.1,][]{Dullemond3}, which provides calculations of continuum radiative transfer in three-dimensional 
axisymmetric (therefore 2-D) circumstellar dust distributions around a central illuminating star. The main radiative transfer calculations are done with a Monte Carlo algorithm. To produce spectra and images, a post-processing with a ray-tracing 
code is done. The disk is described using a passive irradiated flared disk based on the model of \citet{Chiang}, but the code also allows the user to insert a vertical puffed-up inner rim to produce a self-shadowed disk. The space is mapped in spherical 
coordinates. We therefore have radial grid points $r$ and angular grid points $\theta$, where $\theta=0$ is the polar axis and $\theta=\theta_{\rm max}=\frac{\pi}{2}$ is the equatorial plane. We again tried to reproduce the SED and visibilities simultaneously and 
also to fit the silicate absorption feature. To scan the parameter space, more than 10,000 different models were calculated and compared to the observed data. 

The geometry of the disk is described by the pressure scale height $h_{\rm disk}$ at a reference radius $r_{\rm disk}$, and an exponent $p$ in the following way (for $r_{\rm in}<r<r_{\rm disk}$):
\begin{equation}
h(r)= h_{\rm disk}\cdot\left(\frac{r}{r_{\rm disk}}\right)^p.
\end{equation}
Furthermore, either the surface density at the outer radius $\sigma_0$ can be given by the user directly, or (as we did) be calculated from the user-defined total disk mass $m_{\rm disk}= m_{\rm gas} + m_{\rm dust}$. We used a 
gas-to-dust ratio of 100. The surface density as a function of radius is then given by
\begin{equation}
 \sigma (r)= \sigma_0 \cdot \left(\frac{r}{r_{\rm disk}}\right)^{-q}.
\end{equation}
For $r>r_{\rm disk}$, the exponent $q$ is chosen to be equal to 12 such that the density drops so fast that $r_{\rm disk}$ is effectively the outer disk radius. This outer radius was varied between 30~AU and 150~AU. The density distribution 
in the whole disk is described by
\begin{equation}
 \rho (r,\theta) \propto \frac{\sigma (r)}{h(r)\cdot r} \cdot \rm{exp}\left[-\frac{1}{2}\left({\frac{\theta_{\rm max}-\theta}{h(r)}}\right)^2\right].
\end{equation}
Parameters that remained fixed during the modeling process, were the temperature of the central object  $T_{\rm star} =25,000\,\rm{K}$ (according to the spectral type B0-B2), the exponent $q = 12$ for $r > r_{\rm disk}$, 
and the temperature at the inner rim $T_{\rm in} = 1700\,\rm{K}$. For the chemical composition of the dust, the optical properties of astronomical silicate \citep{Draine} and graphite were chosen, using the grain size distribution of the 
MRN model \citep{Mathis}. A mixture of 70\% silicate and 30\% graphite was assumed.

\begin{figure*}
\parbox{18.5cm}{\parbox{9.0cm}{\includegraphics[width=9.0cm, angle=0]{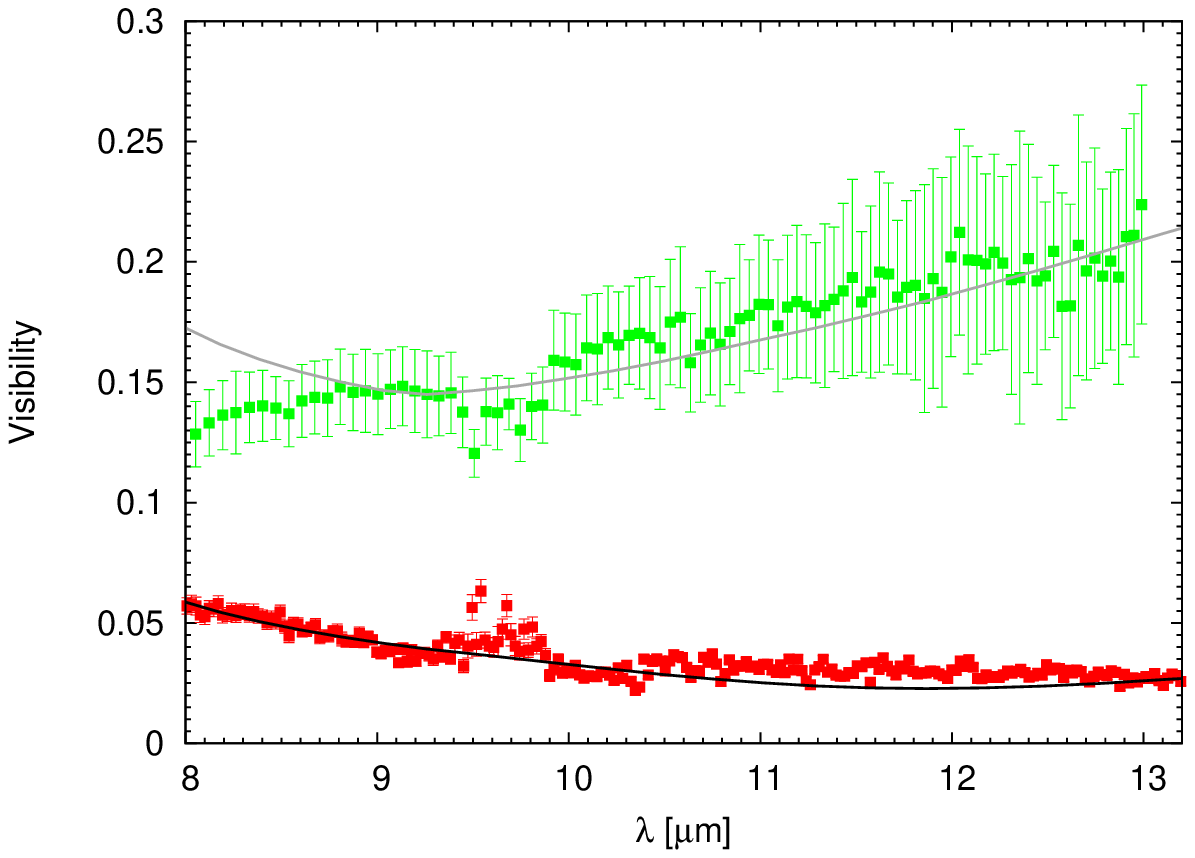}}
\parbox{9.0cm}{\includegraphics[width=9.0cm, angle=0]{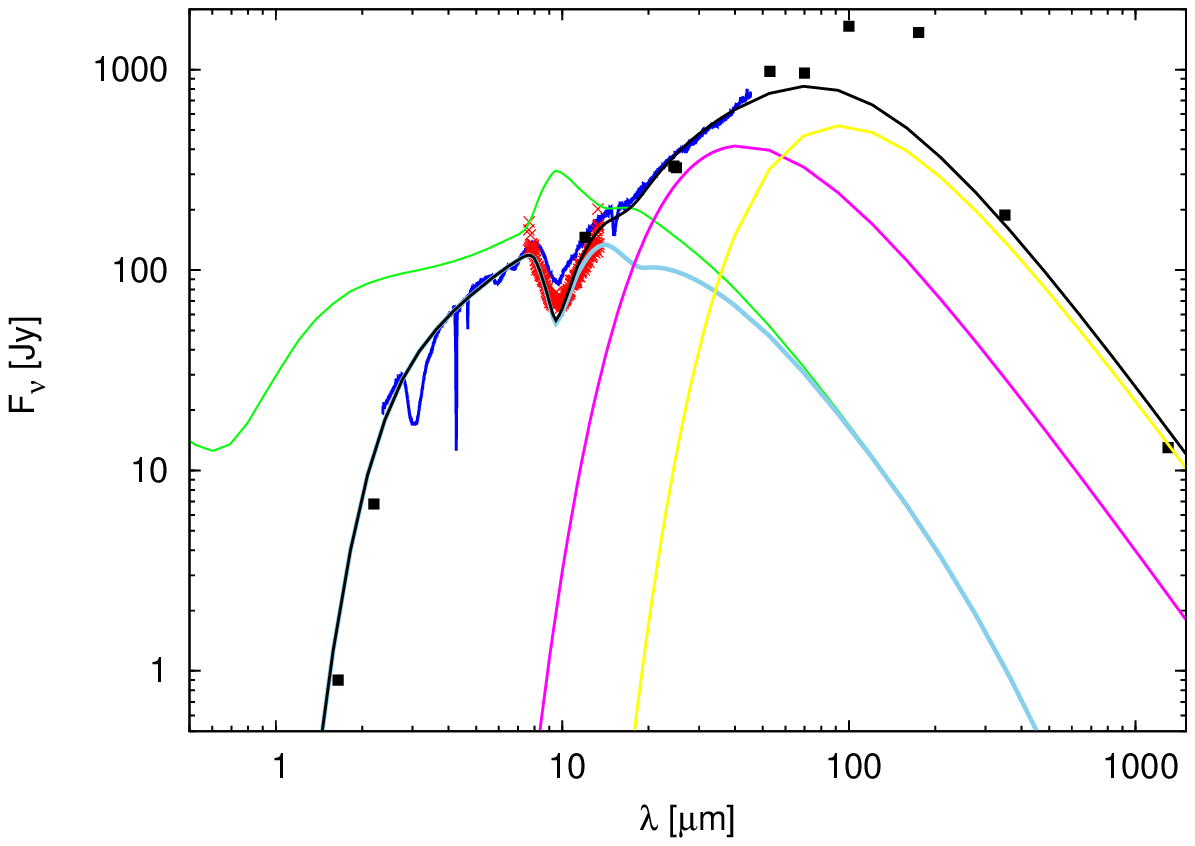} }
\caption{\label{RADMCSpektrum}\textit{Left:} Comparison of the best-fit RADMC model (gray 40~m baseline, black 90~m baseline) for the visibilities of NGC~2264~IRS~1 with the MIDI visibilities (red 90~m baseline, green 40~m baseline). 
\textit{Right:} Comparison of the best-fit RADMC model for the spectrum of NGC~2264~IRS~1 with the ISO spectrum (blue), the MIDI spectrum (red crosses), and the data points from Table~\ref{OtherMeasurements} (black rectangulars). The green 
line is the RADMC model without foreground extinction, the light blue line the model with a foreground extinction of 27~mag, and the black line the RADMC model with extinction and two additional blackbody components with temperatures 
of 55~K (yellow) and 120~K (pink).}}
\end{figure*}

In addition to the original RADMC code, we introduced foreground extinction $A_V$, as otherwise all models failed to reproduce the deep silicate absorption 
feature. Parameters that were varied during the fitting process are shown in the following list together with their values for the best-fit RADMC model:
\begin{itemize}
 \item $r_{\rm disk}$ = 50~AU
 \item $h_{\rm disk}$ = 0.1
 \item $L_{\rm star} = 3.98 \times 10^3\,L_{\sun}$
 \item $p=2/7$
 \item $m_{\rm disk}$ = $0.1\,M_{\odot}$
 \item $q$ = 2.5 for $r < r_{\rm disk}$
 \item $i$ = $30^{\circ}$
 \item PA = $40^{\circ}$
 \item $A_V = 27\,\rm{mag}$.
\end{itemize}
This best-fit model is shown in Fig.~\ref{RADMCSpektrum}. It provides a good fit to the visibilities as well as the SED up to $12\,\mu$m.

The RADMC model describes the central YSO and its circumstellar matter up to the
outer disk radius of 50~AU, i.e.~predicts the emission in the central
100~AU diameter region. Since this area is completely inside the MIDI field-of-view,
 the model visibility can be  directly compared to the observed visibilities.
For the fits to the SED, however, we have to
take into account that the very large beams of the
far-infrared observations must clearly include emission from the
surrounding large-scale cloud, on spatial scales of
 $\sim 20\,000$~AU ($\sim 0.1$~pc)  and larger,
which is far outside the $r=50$~AU model area.
This large-scale cloud emission should not be confused with a possible
circumstellar envelope, but represents the molecular clump and the surrounding cloud
in which IRS~1 is embedded (see discussion below). The mm-maps presented by \citet{Schreyer1} show that the
size of this cloud is about $1'$ (corresponding to $\sim 55\,000$~AU or $\sim 0.27$~pc).

The effects of this large-scale cloud are as follows:
our MIDI measurements are insensitive to this large-scale emission, because it
is far more extended than the angular resolution (and also the field-of-view) of MIDI.
Since the large-scale cloud emission is completely over-resolved for MIDI,
it will not affect the observed visibilities.
The observed fluxes, however, are expected to be strongly affected by the
large-scale cloud. First, the cloud material in front of the embedded object
IRS~1 will cause considerable extinction. This effect is represented by the
foreground extinction we use in our modeling.
The second effect is that the cool dust in the large-scale cloud will
produce far-infrared and mm-emission. Owing to the large beam sizes of the
far-infrared and mm-observations, a large fraction of the observed fluxes
in these beams will result from the large-scale cloud, and not from the central embedded
object IRS~1. This explains why our RADMC model fluxes at wavelengths $\ga 20\,\mu$m
are considerably lower than the observed fluxes.
To approximately include this large-scale cloud emission component
in our model, we added to our SED model in Fig.~\ref{RADMCSpektrum} two additional blackbody
components. For the temperature of the first component, we used $T=55$~K, as
determined by \citet{Schreyer1} for the clump in which IRS~1 is embedded.
For the second component, we used a temperature of 120~K, which is typical
of warm dust around massive YSOs in so-called ``hot-cores'' \citep{Herbst}. Fig~\ref{RADMCSpektrum} shows that with the addition of these two components good agreement
between the observed fluxes and the model can be achieved.

An interesting question is whether this 120~K component could be considered
as an envelope around the disk, as for one interpretation of
hot cores as an infalling envelope undergoing an intense accretion phase \citep{Osorio}.
The RADMC model of a disk without surrounding envelope provides a very good fit of both visibilities.
Adding a spherical circumstellar envelope with a radius of $\leq 250$~AU would
change the model visibilities significantly (by $> 5\%$) and lead to disagreement between
the model and the data.
On the other hand, a (more or less) homogeneous envelope with a radius of $> 250$~AU would be
over-resolved for MIDI and not affect the visibilities.
However, to reproduce the far-infrared fluxes in the SED,
the radius of the 120~K emitting region has to be at least
$\geq 400$~AU (in the limiting case of optically thick blackbody emission). Owing to the limited MIDI field of view,
the difference between the ISO and the MIDI spectrum could result from such an extended
structure. The question of whether an extended circumstellar envelope is present 
can thus not be answered by the available data. Whether the material in such a
hypothetical envelope would be gravitationally bound and ultimately accrete
onto the circumstellar disk, remains another open question.

\section{Discussion\label{Discussion}}
We now want to discuss our findings presented in Sect.~\ref{Modelling}. We first compare the size of the mid-IR emission region as calculated in Sect.~\ref{Geometry} to the corresponding sizes of similar objects. We then discuss the radiative 
transfer models. 
\subsection{Comparison to MIR sizes of other massive YSOs}
The near-IR sizes of the disks of most YSOs (in particular Herbig~Ae stars) show a tight correlation with the luminosity \citep{Monnier1} because the emission in the NIR comes from regions close to the dust sublimation radius, which is sensitive 
to radiation from the central star and scales approximately with the square root of the luminosity. The mid-infrared emission, in contrast, originates from a wider range of regions: the hot inner wall located at the dust sublimation radius, the surface 
of an irradiated disk, and the circumstellar envelope. As shown in \citet{Monnier}, the size of the region emitting in the MIR therefore spans a much larger region for a given luminosity. For circumstellar disks, factors such as the 
flaring of the disk and the shadowing caused by the inner rim play an important role. Here, we wish to compare the MIR size of NGC~2264~IRS~1 as found for the uniform disk model with the MIR sizes of other MYSOs, as well as with those of Herbig Be stars. 

\begin{figure*}
\parbox{18.5cm}{\parbox{9.0cm}{\includegraphics[width=9.0cm, angle=0]{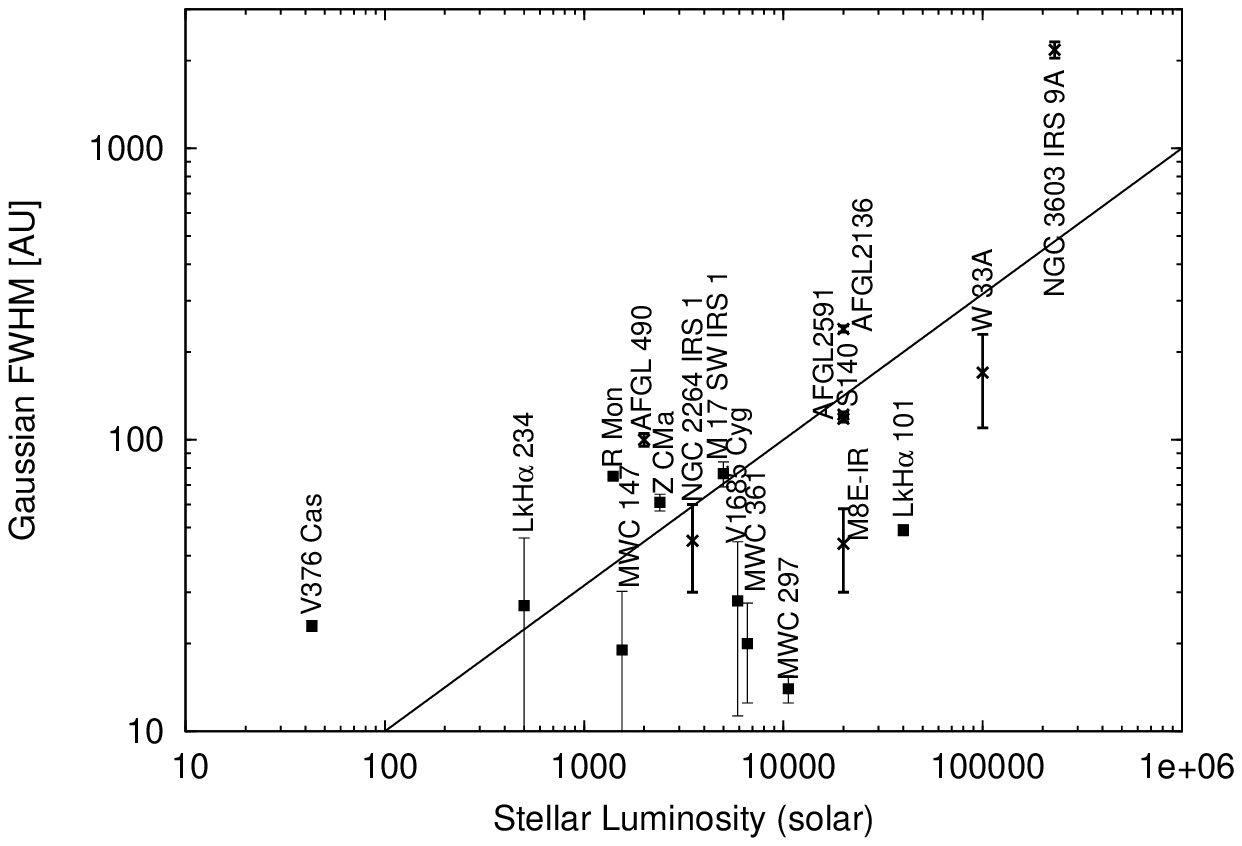}}
\parbox{9.0cm}{\includegraphics[width=9.0cm, angle=0]{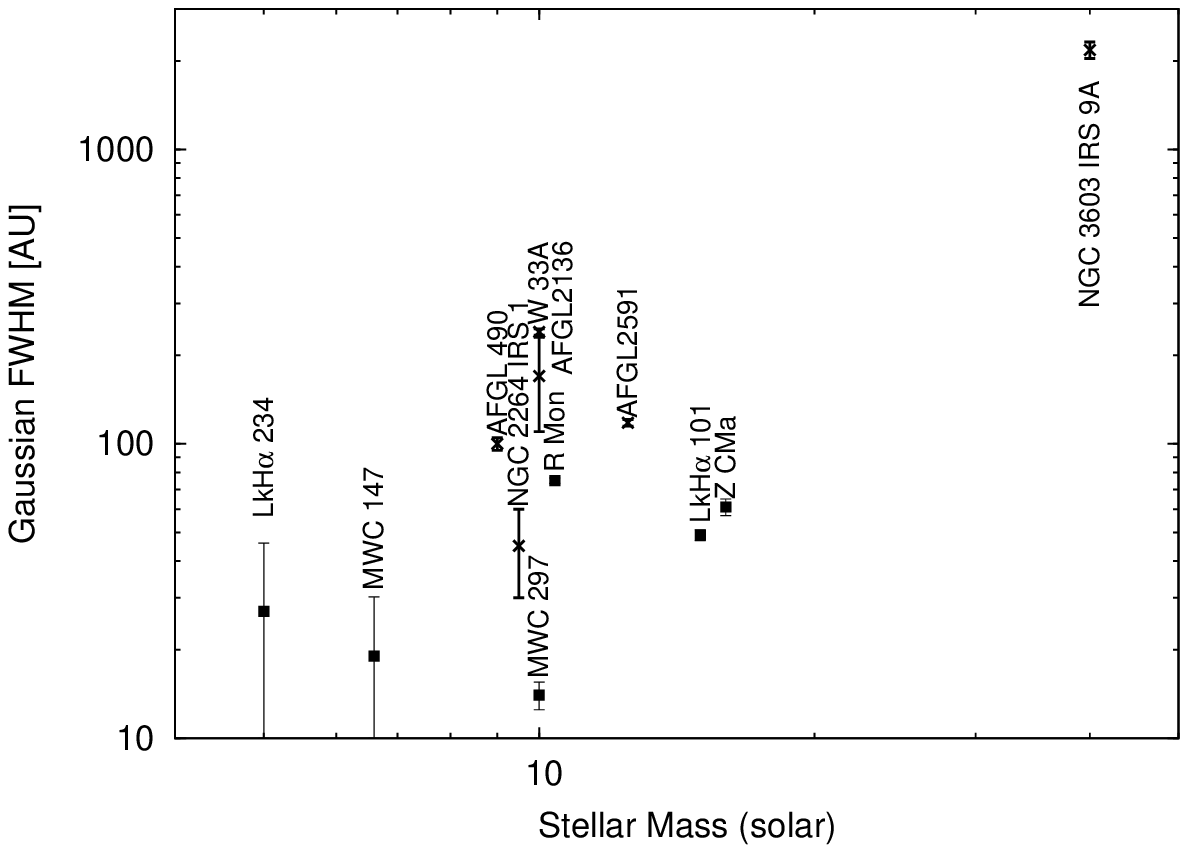}}
\caption{ \label{Sizes}\textit{Left:} Mid-infrared-size-luminosity diagram. Rectangulars correspond to Herbig~Be stars, crosses to MYSOs. Values and references are shown in Table~\ref{MIRsizes}. The black line corresponds to a $R\propto\sqrt{L}$ relation.
 \textit{Right:} Mid-infrared-size-stellar-mass diagram. Rectangulars correspond to Herbig~Be stars, crosses to MYSOs.}}
\end{figure*}

Only a few other MYSOs have been studied with high angular resolution in the MIR until now. Amongst them are AFGL~2591, AFGL~2136, AFGL~490, and S140~IRS~1 from the study of MIR sizes of \citet{Monnier}, as well as  W~33A \citep{deWit1,deWit2}, 
and NGC~3603-IRS~9A \citep{Vehoff}. The sample of Herbig Be stars was also taken from \citet{Monnier}. The disk size vs. luminosity diagram of these objects and NGC~2264~IRS~1, as well as the size vs. stellar mass diagram is shown in 
Fig.~\ref{Sizes}. Parameters such as stellar mass, distance, and the numbers for disk size and luminosity can be found in Table~\ref{MIRsizes}. Both luminosity and mass of the central object seem to be correlated with the size of the MIR emission 
region. Owing to the factors considered above, objects with luminosities ranging between 1000 and 10,000\,$L_{\odot}$ display a large scatter in their sizes (a factor of around 10). The same is true for the mass range of~$\sim 8-20$\,$M_{\odot}$. 
The size of the MIR emitting region around NGC~2264~IRS~1 seems to be typical for their mass and luminosity. 

\subsection{Discussion of the radiative transfer models} 
Models of a circumstellar envelope only did not provide good fits, as they were unable to reproduce visibilities and NIR flux simultaneously. Furthermore, the estimated sizes of the dust distribution (Sect.~\ref{Geometry}) are different for 
different position angles, which implies that there is an asymmetric dust distribution.

The radiative transfer modeling performed with RADMC suggests that IRS~1 has an optical thick, but geometrically thin, flaring circumstellar disk with a mass of $0.1\,M_{\odot}$. This is supported by the results of the SED fitting process using the 
Robitaille grid. Not only the best-fit Robitaille disk model but also the ten best-fit models have parameters describing the central 
object and the circumstellar disk that compare very well with those of the best-fit model found using RADMC. However, the inclinations of the Robitaille models vary between $20^{\circ}$ and $70 ^{\circ}$. The temperature and mass of the central object are slightly 
higher in all of these disk models than the values used for the RADMC model.

\begin{figure}
\includegraphics[width=9.0cm, angle=90]{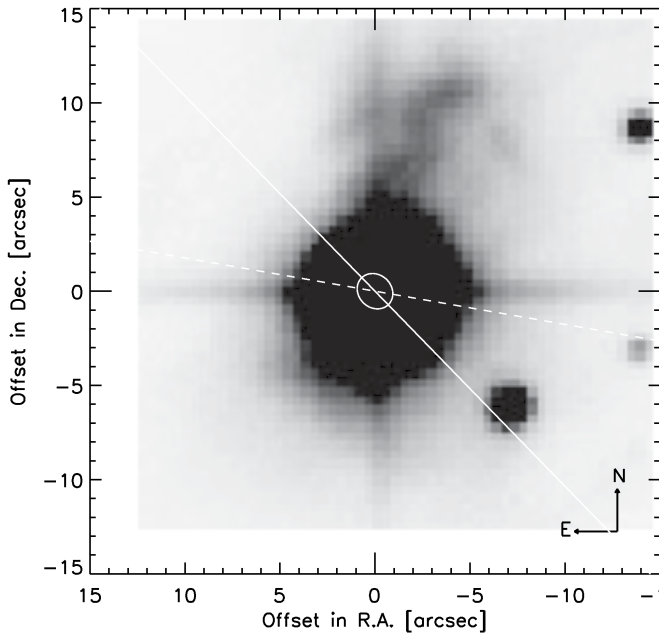}
\caption{ \label{Jet}K-band image of NGC~2264~IRS~1 from \citet{Schreyer1}. We overlay an ellipse (white) that shows a disk seen under an inclination of $30^{\circ}$ and a position angle of $40^{\circ}$. The size of the disk is not drawn to scale. 
The solid white line shows the orientation of the 40~m baseline, the dashed one the orientation of the 90~m baseline.}
\end{figure}

The two blackbody components producing the far-IR flux for our best-fit RADMC model could originate from a large surrounding cloud or envelope, respectively, that would with the large angular resolution of MIDI only be seen as foreground extinction. 
This would agree with the large foreground extinction ($> 20$~mag) as used in the best-fit RADMC and derived from the ten best-fit Robitaille disk models. Furthermore, previous publications have suggested that there is a large amount of 
foreground extinction with values between 
20~mag \citep{Thompson} and 35~mag \citep{Tokunaga}. The inclination of the best-fit RADMC model is $30^{\circ}$, but owing to the poor uv-coverage this value is poorly constrained. The position angle of the jet-like feature seen in the near-IR 
is $20^{\circ}$ (see  Fig.~\ref{Jet}), and the position angle in our best-fit disk model is 40$^{\circ}$. This does not match perfectly, but the value for the 
position angle is also poorly constrained. However, an overall geometric picture of a moderately inclined disk seems reasonable. 

\section{Summary \& conclusions\label{Summary}}
We have presented MIDI mid-infrared interferometric observations of the massive young stellar object NGC~2264~IRS~1. The observed visibilities provide \textit{no} hint of multiplicity of the central source in the separation range 
from $\sim 30$~AU (resolution limit) to $\sim 230$~AU (slit width limit). Geometric models (Sect.~\ref{Geometry}) suggest a size of $\sim 30-60$~AU for the mid-infrared emission region. This value is in good agreement with the mid-infrared sizes 
of similar objects. 

Our modeling shows that a simple temperature-gradient model for a circumstellar disk can reproduce the SED at mid- and far-IR wavelengths, but fails to reproduce the SED in the near-IR and the observed visibilities; this model 
can therefore be ruled out.

We have performed a detailed radiative transfer modeling of the observed SED and visibilities with the RADMC code. It suggests a scenario of a geometrically flat but optically thick circumstellar dust disk. We also used the fitting tool 
by \citet{Robitaille}, which leads to similar disk models and therefore confirms the results obtained with RADMC, whereas envelope models found within the Robitaille grid did not 
provide similarly good fits to SED and visibilities simultaneously. Although there is no observational evidence, the possible presence of a large ($r \geq 400$~AU) spherical envelope around the disk cannot be excluded by the data. 
The mass of the circumstellar disk is about $0.1\,M_\odot$. The data suggest an inclination angle of $\sim 30\degr$ and a position angle 
of $\sim 40\degr$ for the orientation of the disk. These values are consistent with an overall geometrical model based on the jet-like feature seen to the north-east of IRS~1. Comparing NGC~2264~IRS~1 to other YSOs, the size of its MIR emitting region 
seems to be typical of its luminosity and mass. However, the large scatter of sizes in this range of luminosities and masses points towards a wide variety of (disk) morphologies among these objects. 

More observational research, such as future MIR 
interferometric observations, will provide tighter constraints on the circumstellar material around this interesting source and therefore will help us to  clarify our understanding of the disks around high-mass YSOs.

\begin{acknowledgement}
We would like to thank the anonymous referee for his suggestions and comments that helped to improve the paper. We would also like to thank K.~Dullemond for explanations and discussions about RADMC.
We gratefully acknowledge funding of this work by the German \textit{Deutsche Forschungsgemeinschaft, DFG} project number PR~569/8-1. 
\end{acknowledgement}

\bibliographystyle{aa}
\bibliography{Paper}

\appendix
\section{Parameters and references for size-luminosity plot}
\begin{table*}
\caption{\label{MIRsizes}MIR-sizes of MYSOs and Herbig Be stars}
\centering
 \begin{tabular}{c c c c c c c}
 \hline\hline
 Object & Diameter & Distance & Luminosity    & Mass          & SpT & Reference \\ 
  {}	& [AU]	   & [pc]     & [$L_{\odot}$] &	[$M_{\odot}$] & {}  & {} \\ \hline
 \object{NGC 3603 IRS 9A}& 2177 		& 7000 	& 2.3$\cdot 10^5$	& 40  	& {} 	& 1 \\ 
 \object{W 33A}		& 115-230	& 3800	& 1$\cdot 10^5$		& 10	& {} 	& 2 \\ 
 \object{V376 Cas}	& 23		& 630	& 43			& {}	& B5e	& 3, 4\\ 
 \object{MWC 147}	& 19		& 800	& 1550			& 6.6	& B6	& 3, 5, 6\\ 
 \object{R Mon}		& 75		& 800	& 1400			& 10.4	& B0	& 3, 7, 8\\ 
 \object{Z CMa}		& 61		& 1000	& 2400			& 16	& B	& 3, 9, 10\\ 
 \object{MWC 297}	& 14		& 250	& 10600			& 10	& B1.5	& 3, 11, 12\\ 
 \object{V1685 Cyg}	& 28		& 980	& 5890			& {}	& B3	& 3, 4\\ 
 \object{MWC 361}	& 20		& 440	& 6600			& {}	& B2	& 3, 4\\ 
 \object{LkH$\alpha$ 234}& 27		& 1000	& 500			& 5	& B7	& 3, 13, 14\\
 \object{AFGL 490}	& 100		& 1000	&2$\cdot 10^3$		& 8-10	& B2	& 3, 15\\ 
 \object{LkH$\alpha$ 101}& 49		& 700	& 2-6$\cdot 10^4$	& 15	& Be	& 3, 16\\
 \object{S140 IRS 1}	& 122		& 910	& 2$\cdot 10^4$		& {}	& {}	& 3, 17, 18\\ 
 \object{AFGL 2136}	& 240		& 2000	& 2$\cdot 10^4$		& $>$10	& {}	& 3, 19\\ 
 \object{AFGL 2591}	& 118		& 1000	& 2$\cdot 10^4$		& 10-15	& {}	& 3, 20\\ 
 \object{M 17 SW IRS 1}	& 69-84		& 2100	& 5$\cdot 10^3$		& {}	& B0	& 21 \\ 
 \object{M8E-IR}		& 30-57		& 1500	& 2$\cdot 10^4$		& {}	& B0	& 22 \\ 
 \object{NGC 2264 IRS 1}	& 30-60		& 913	& 1.5-4.7$\cdot 10^3$	& 9.5	& B0-B2	& see Chapt.~\ref{Object}, \ref{Geometry}\\ \hline
 \end{tabular} 
\tablebib{(1)~\citet{Vehoff}; (2)~\citet{deWit1}; (3)~\citet{Monnier}; (4)~\citet{Acke1};
(5)~\citet{Kraus2}; (6)~\citet{Hernandez}; (7)~\citet{Murakawa1}; (8)~\citet{Murakawa2};
(9)~\citet{Alonso}; (10)~\citet{Thiebaut}; (11)~\citet{Acke2}; (12)~\citet{Drew}; (13)~\citet{Shevchenko}; (14)~\citet{Fuente}; (15)~\citet{Schreyer3}; (16)~\citet{Herbig}; (17)~\citet{Crampton}; (18)~\citet{Lester}; (19)~\citet{Kastner1}; 
(20)~\citet{Tak}; (21)~\citet{Follert}; (22)~\citet{Linz}.}
\end{table*}
\end{document}